\pdfoutput=1

\documentclass[11pt]{article}

\usepackage[final]{acl}

\usepackage{times}
\usepackage{latexsym}

\usepackage[T1]{fontenc}

\usepackage[utf8]{inputenc}

\usepackage{microtype}

\usepackage{inconsolata}

\usepackage{graphicx}

\usepackage{multirow}
\usepackage{caption}
\usepackage{subcaption}
\usepackage{booktabs}

\usepackage{xspace}

\newcommand{\noxai}{no-XAI\xspace}
\newcommand{\lime}{LIME\xspace}
\newcommand{\chatgpt}{GenAI\xspace}

\newcommand\uw{$^\spadesuit$}

\newcommand\ai{$^{\diamondsuit}$}
\newcommand\aspace{\hspace{.75em}}

\graphicspath{{./src/img/}}

\title{How Performance Pressure Influences AI-Assisted Decision Making}

\author{
  Nikita Haduong\uw\aspace
  Noah A. Smith\uw\ai\aspace
  \\
  \uw{}University of Washington \aspace
  \ai{}Allen Institute for AI\\
  {\tt \{qu,nasmith\}@cs.washington.edu 
  }
}

\begin{document}
\maketitle
\begin{abstract}
Many domains now employ AI-based decision-making aids, and although the potential for AI systems to assist with decision making is much discussed, human-AI collaboration often underperforms due to factors such as (mis)trust in the AI system and beliefs about AI being incapable of completing subjective tasks.  One potential tool for influencing human decision making is
performance pressure, which hasn't been much studied in interaction with human-AI decision making.
In this work, we examine how pressure and explainable AI (XAI) techniques interact with AI advice-taking behavior. 
Using an inherently low-stakes task (spam review classification), we demonstrate effective and simple methods to apply pressure and influence human AI advice-taking behavior by manipulating financial incentives and imposing time limits.
Our results show complex interaction effects, with different combinations of pressure and XAI techniques either improving or worsening AI advice taking behavior.
We conclude by discussing the implications of these interactions, strategies to effectively use pressure,
and encourage future research to incorporate pressure analysis.
\end{abstract}

\section{Introduction}
With modern language models facilitating interaction with various AI systems, decision aids are now available across many industries (e.g., medical  diagnoses \citep{dilsizian2014artificial,duron2021assessment}; financial management \citep{zopounidis2002multi}; and criminal recidivism risk, \citealp{mckay2020predicting}), and when used to complement human abilities, have the potential to outperform either the human or AI working alone.
The potential is not necessarily realized, however, because of several challenges: debates on ethical resposibility of decisions \citep{smith2021clinical,busuioc2021accountable,johnson2021algorithmic}, the human ability to recognize when AI advice should be taken \citep{10.1145/3581641.3584066}, 
mental models (biases) regarding AI performance and ability  to perform well on subjective tasks \citep{clark-etal-2021-thats,jones2023people}, and effects of how the AI advice is delivered ~\citep{steyvers2023three}.

Many research directions thus aim to resolve these barriers to complementarity in human-AI performance, measured as \textit{appropriate AI advice reliance}, which has two dimensions: taking correct AI advice and disregarding incorrect AI advice. Investigations include the effects of showing explanations from explainable AI (XAI) alongside AI system predictions \citep{10.1145/3411764.3445717}, introducing cognitive forcing functions when presenting AI advice \citep{buccinca2021trust}, adjusting AI advice presentation methods \citep{rastogi2022deciding}, and adjusting task framing to account for biases about the types of tasks AI can do \citep{doi:10.1177/0022243719851788}. 

In AI-assisted decision making, the human makes the final decision, bearing full responsibility for its consequences. It is established that performance pressure from responsibility can influence decision making behavior \citep{ashton1990pressure}. But how does it influence \textit{AI-assisted} decision making?
AI-assisted decision making experiments have considered tasks with stakes that are intrinsically high (loan defaults; \citealp{10.1145/3359152}) and low (speed dating; \citealp{doi:10.1177/0022243719851788}), but the stakes have little tangible effect or implication for evaluators. Hence, we observe a gap in the literature of how people rely on AI assistants \textit{under performance pressure}, that is, when stakes matter personally.

We believe this question is of special significance in the NLP research community, and not only in deployment scenarios.
Modern NLP  relies on eliciting high-quality data from humans to train models, often with systems in the loop.  For example, Dynabench \citep{kiela-etal-2021-dynabench} and ANLI \citep{nie-etal-2020-adversarial} are datasets where humans work with AI models to create data through finding adversarial or interesting examples.  Such datasets are often curated with low personal stakes, e.g., \citet{wadhwa-etal-2024-investigating}, \citet{krishna-etal-2023-usb,krishna2024genaudit}, \citet{lu2024does} and \citet{haduong-etal-2023-risks} crowdsourced annotations and paid hourly wages. Could judiciously applied performance pressure influence the decisions of annotators building research datasets in ways that lead to improvements in data, and by extension AI evaluations and systems?

In this work, we seek to understand how performance pressure influences AI advice usage when the advice is provided as a second opinion.
We recruit participants to decide whether a hotel review is genuine or deceptive and provide them with an AI advisor.
We manipulate performance pressure in three different ways: by providing a bonus for correct answers, by deducting from the task compensation for incorrect answers, and by providing a bonus for correct answers within a time limit.
We further investigate how performance pressure and different XAI techniques interact.
Our results reveal a complex story. Under certain conditions, pressure can either improve or lower appropriate AI advice reliance, and XAI can sometimes mitigate negative effects of pressure.

Our contributions are:
\begin{itemize}
    \item We demonstrate how to increase the stakes in an inherently low-stakes AI-assisted decision making setting; this approach can generalize to many pre-existing study designs. 
    \item We show how XAI affects advice reliance (both positively and negatively) and interacts with pressure, forming a complex picture about how human behavior changes. These findings suggest opportunities for designing adaptive decision-making environments when different XAI methods are available.
    \item We explore how pressure and confirmation bias can increase overreliance on AI advice and discuss implications of unintentionally encouraging people to trust AI \emph{too} much. 
\end{itemize}

\section{Related Work}

Designing systems that can assist with decision making is difficult, because people are influenced by many factors when taking advice, such as their personal expertise \citep{ronayne2019ignoring}, the advisor's reputation \citep{yaniv2000advice}, or the style of advice delivery (e.g., inviting or broadcasting, \citealp{CHHABRA2013573,morrison2024impact}), resulting in inconsistent advice taking behavior that can be challenging to predict. For example, even if advice is objectively high-quality (e.g., advice based on fact), it may still be disregarded \citep{wang2018does}.
AI advisors should complement human decision making to achieve higher collaborative performance, compared to individual performance, but recent studies have observed that over- and under-reliance on AI advice result in suboptimal collaboration  \citep{7349687,21e0babd97564ce69ad46fe4944fae94}.

\paragraph{Toward appropriate AI advice use.} Algorithmic aversion has been shown to be task-dependent, in line with ideas about how well machines can perform on subjective tasks. When the task is subjective, e.g., predicting  speed dating results, \citet{doi:10.1177/0022243719851788} found increased algorithmic aversion, as opposed to an objective task, e.g., predicting financial outcomes.
Hypothesizing that people discount AI advice because they do not trust the AI system, researchers have used explainable AI (XAI) methods and shown the explanations alongside the AI advice. Many studies have observed XAI positively influencing AI reliance (e.g., \citealp{panigutti2022understanding,ben2021explainable,lee2023limeade}).
Yet \citet{fleis-algo-aversion} observed the opposite: when decisions were about quantifiable skills (e.g., work experience or command of English), rather than soft skills (e.g., diligence or ability to work in teams), adding explanations did not significantly increase AI advice reliance.
\citet{jiang2022needs} similarly observed XAI failing when the user is too uncertain.
Another set of methods aim to mitigate inappropriate AI reliance through cognitive forcing functions---interventions that cause a decision maker to engage in analytical thinking 
\citep{cog-forcing-function}.  For example, \citet{rastogi2022deciding} successfully employ a cognitive forcing function to reduce anchoring bias---a bias where people weight earlier information higher---by adding a time delay before showing AI advice.

\paragraph{Decision making under pressure.}
An important environmental factor to consider is the influence of stressors on the human decision maker. 
Decision making often occurs under time stress or the weight of responsibility, for example.
Different stressors can influence decision making in different ways \citep{HENDERSON2024102592}, and when multiple stressors are present, their compound effect can present itself in additive, synergistic, or antagonistic ways \citep{https://doi.org/10.1002/ece3.2609}.
The influence of stress on AI-assisted decision making is an understudied factor, although in recent work,
\citet{time-pressure} study how AI-assisted decision makers perform under \emph{time} pressure, which emerges in real-world settings like operating rooms and search and rescue missions. They study \emph{when} to provide AI advice in an inherently high-stakes medical diagnosis task, adapted to be approachable to laypeople.
\citet{gazit2023choosing} studied AI-assisted decision making under the pressure of responsibility and observed how responsibility pressures overrode logical reasoning, resulting in lower appropriate AI reliance.
The experiment setup involved surveying managers in business organizations, using experts with real responsibilities but asking about their behaviors rather than empirically observing them.
Further work is needed to understand the role of responsibility and pressure in AI-assisted decision making.

\paragraph{Manipulating performance pressure.}
Performance pressure can be experimentally manipulated through different consequences, e.g., rewards and reputation \citep{stoker2019effect}.
High-quality crowdworker data can be collected by using an appropriate financial incentive in the form of a fair base pay and bonuses.
A higher potential reward, or bonus, can increase the pressure on the crowdworker toward higher performance.
A common way of presenting the bonus is to frame it as a {\em gain}, e.g., ``if you do a good job, you can earn a bonus''.
Alternatively, the bonus could be framed as a {\em loss}, e.g., ``if you do a poor job, you will lose your bonus''. 
The literature in risk aversion, the propensity to play it safe, and loss aversion, the fear of losing out, has observed a stronger pressure effect from framing incentives as a loss rather than a gain \citep{merriman-loss-aversion}.
\citet{grgic2022taking} designed a study investigating how trust in the AI advisor evolves and successfully used the loss framing. In their experiment, users made AI-assisted decisions and updated their mental models of the AI behavior.
Here, we are interested in studying the influence of external performance pressure stressors to encourage \emph{more appropriate AI reliance} (i.e., correctly using AI advice to improve decision-making), rather than studying effects on user trust. 

\section{Experiment} \label{sec:exp1}

From that extant literature, we form the following hypotheses:

{\bf H1}: We can influence AI advice reliance by manipulating the environmental pressure. Increased performance pressure from monetary incentives framed as a loss will improve appropriate AI advice reliance, and increased pressure from time limits will reduce it (e.g., \citealp{Zakay1993}).

{\bf H2}. The risk aversion level of participants and their trust in the AI advisor can predict the influence of performance pressure (e.g., \citealp{chokingTimePressure}). Participants with higher risk aversion will be more careful in their decision making. Participants with higher trust in the AI advisor will have more decisions aligned with the AI advice.

{\bf H3}. Performance pressure will act as a cognitive forcing function, influencing participants to spend more time when making their decisions, because they want to be more careful about their response. 

{\bf H4.} The positive effects from XAI will hold under pressure, potentially further increasing appropriate reliance of AI advice over no XAI.

To study how pressure influences AI advice-taking with and without AI explanation aids, we recruit Prolific\footnote{\url{https://www.prolific.com/}} crowdworkers and task them with judging whether a hotel review is genuine or deceptive. We design a within-subjects experiment manipulating environmental pressure and run three experimental settings simultaneously, changing the availability of an explanation aid, to consider all of the above hypotheses at the same time.

\subsection{Dataset}
Of the many text annotation tasks available, we choose deceptive review classification because it has real-world importance, is not an inherently high-risk task (as compared to medical diagnosis), does not require expertise in the real-world setting (as compared to criminal recidivism), and likely has minimal relevance to our participants (e.g., the impact of predicting a review incorrectly has no personal effect on the participant). Hence, the pressure to perform well on this task must be primarily be external, necessitated by our experimental setup where we wish to simulate different levels of external performance pressure.  It also parallels the annotation setting of data creation for NLP research.

We draw data from the Deceptive Opinion Spam Corpus \citep{ott-etal-2011-finding,ott-etal-2013-negative}, a binary classification task, which contains genuine hotel reviews from travel websites and deceptive reviews written by Amazon Mechanical Turk workers. 
The task is challenging: human performance is 55\%---little better than random chance, ensuring that   AI advice taking behavior we observe is not confounded by participants' prior knowledge or skill. 

\subsection{XAI Methods}\label{sec:xai}
We use two XAI methods: feature importance highlighting (\lime; \citealp{lime}), and natural language explanations produced by a generative AI (\chatgpt).  \lime requires feature weights, thus we train an SVM classifier with tf-idf features. Our model achieves 86\% accuracy on the test set using 5-fold cross-validation, in line with the SVM used by \citet{10.1145/3581641.3584066}.
We do not disclose the model performance in our study to avoid user bias about objective performance metrics of the advice.
For \chatgpt, we generated the explanation by prompting a large language model, ChatGPT\footnote{Accessed January 20, 2025}, to explain why a review received a particular label (Appendix \ref{app:prompt-chatgpt}). Note that this explanation is hallucinated, and the same review could receive a generated explanation for either label. 
We selected these approaches because 
\lime and \chatgpt are popular XAI methods used for text and studying AI-assisted decision making \citep{10.1145/3581641.3584066,10.1145/3411764.3445717}.\footnote{We have no hypotheses about different kinds of AI methods or their accuracy, nor about the faithfulness of the XAI explanations to the workings of the classifier.  Hence we opted for a relatively simple but realistic classification system and widely-used XAI methods.}

\subsection{User Interaction}
To measure the influence of AI advice under different stakes, we require a sequential decision making setup. For this reason, we use the judge-advisor system (JAS; \citealp{sniezek1995cueing}). Under JAS, a user will first make a judgment alone, then receive advice, and finally make a second judgment (either confirming or adjusting their initial judgment). The sequential nature allows us to measure influence by comparing the final judgment with initial pre-advice judgments. 
Our interface design is heavily inspired by \citet{10.1145/3581641.3584066} in order to establish our baseline 
conditions with previous work.

\subsection{Independent Variables. }
We measure demographics data (gender, education level, race), trust in AI and frequency of AI usage in work (5-point Likert), trust in the AI advisor (4 items), and risk aversion (Appendix \ref{presurvey}). Risk aversion is measured in two ways: the 10-item IPIP representation of the \citet{mpq} Multidimensional Personality Questionnaire\footnote{\url{https://ipip.ori.org/newMPQKey.htm}} (MPQ; Appendix \ref{mpq-questions}) and the \citet{holt-laury}  Risk Assessment (HL; Appendix \ref{hl-survey}) (10 items). MPQ asks subjects to  rate their level of agreement with statements (e.g., ``I avoid dangerous situations'') using a 4-point Likert. HL contains a list of ``gambles'' where participants choose between ``safe'' and ``risky'' choices. 
Users are incentivized to answer truthfully on the survey to earn a bonus of up to 3.85USD.

\subsection{Experiment Setup}

\paragraph{Items.} We sample 24 reviews from the test set, ensuring a balanced sample of genuine and deceptive reviews, and also of correct vs.~incorrect AI predictions. All reviews had positive polarity. We select two additional reviews for practice: one where the AI is correct and one where it is incorrect. The reviews had a length of 45--120 words.
Each pressure condition (details below) was assigned a balanced, random assignment of 8 reviews (2 of each \{genuine, deceptive\} $\times$ \{correct, incorrect\}), and participants encountered pressure conditions in random order. We included two attention checks and rejected data from participants who failed both.

\paragraph{Subjects.} A total of 302 participants were recruited on Prolific across three explainable AI (XAI) conditions. The recruitment conditions were 95\% HIT acceptance rate, native English speaker, and limited to U.S. workers. After subjects accepted the task, they were directed to a consent form, completed a presurvey with demographics questions, questions about AI usage frequency and trust, and MPQ, 
then received instructions for the task. They completed two practice items and received feedback on the correctness of their decision to ensure they understood the JAS setup and also that the AI advice could be incorrect. For each item, reviewers decided whether a review was genuine or deceptive and rated their decision confidence on a 7-point Likert scale, then received AI advice, then were given the chance to update or confirm their decision and confidence level. Participants did not receive feedback on the correctness of their judgments after the practice items, to ensure that trust in the AI system was held constant across items. After two practice tasks, which are excluded from analysis, subjects judged all 24 
reviews, then completed a postsurvey. The postsurvey contained questions to determine the level of trust in the AI advisor and HL. 
The study was approved by our institution's IRB, and participants were guaranteed a wage of 20USD/hr. Overall, the study took about 30 minutes per participant.
The participants received 6USD base pay for completing the task and were aware of the bonuses. The average wage rate after bonuses was 33USD/hr. Participants whose performance resulted in underpayment received bonuses to meet the wage rate.

\paragraph{Experimental conditions.} 
We use a within-subjects design varying the type of pressure (baseline pressure, payment pressure, or time pressure).
In the baseline pressure condition, participants are informed they will receive a bonus of 0.5USD for every correct decision.
In the payment pressure condition, participants are additionally informed they will lose 0.8USD for every incorrect answer.
In the time pressure condition, participants must make a correct decision within 30 seconds to receive the 0.5USD bonus. A  timer was displayed to indicate remaining time. If participants ran out of time, the timer would count down negatively.
We also use a between-subjects design to study the effects of XAI decision aids (\noxai, \lime, or \chatgpt). The baseline (\noxai) subject group received no explanation for the AI's prediction, the \lime subject group received a LIME feature importance explanation (where text is highlighted to indicate its association with a label), and the \chatgpt subject group  received a natural language explanation generated by ChatGPT (see \S\ref{sec:xai}).

Appendix Fig. \ref*{fig:user-interface-xai} shows an example of the interface with  payment pressure and \lime explanations.

\paragraph{Dependent variables.}  \label{sec:dvs}
Following \citet{10.1145/3581641.3584066}, we measure AI reliance through relative positive AI reliance (RAIR; Appendix A Eq.~\ref*{math:rair}), defined as the ratio of the number of cases where the human relies on AI advice to correct their decision (i.e., they were incorrect before receiving advice and correct after), and relative positive self-reliance (RSR; Appendix A Eq.~\ref*{math:rsr}), defined as the ratio of the number of cases where the human correctly maintains their judgment, disregarding the incorrect AI advice.\footnote{An alternative measure frequently used is weight on advice \citep{10.1145/3686164,schmitt2021towards,harvey1997taking,logg2019algorithm,mucha2021interfaces}.} We also measure overall task accuracy. For all three measures, higher values are preferable.

\begin{figure*}[t]
    \centering
    \begin{subfigure}{.45\textwidth}
        \centering
        \includegraphics*[width=\textwidth]{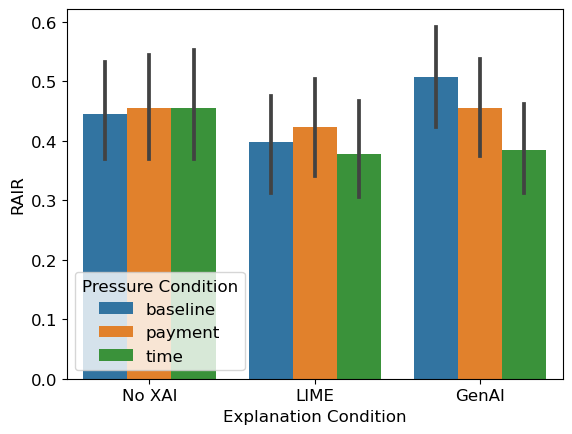}
        \caption{RAIR:  higher indicates less underreliance.}
        \label{fig:rair}
    \end{subfigure}
    \begin{subfigure}{.45\textwidth}
        \centering
        \includegraphics*[width=\textwidth]{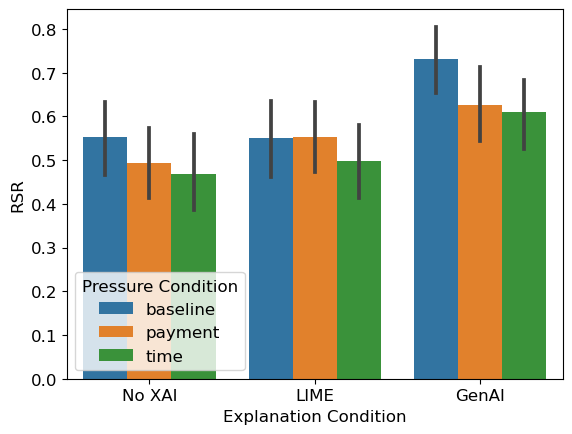}
        \caption{RSR: higher indicates less overreliance.}
        \label{fig:rsr}
    \end{subfigure}

    \caption{GenAI improves appropriate AI advice reliance, but pressure has a predominantly negative effect.}
    \label{fig:rair-rsr}
\end{figure*}

\section{Results}

After filtering for failed attention checks, we collected responses from 99, 102, and 101 subjects across the three XAI conditions: baseline, \lime, and \chatgpt, respectively. Details for how significance testing is conducted are in Appendix \ref{sec:sigtests}, and uncorrected p-values are reported in the following section where relevant.

\paragraph{Demographics and surveys.} 
Participants' demographics reported were: 51\% male, 47\% female, and 2\% other; aged 18--30 (34\%), 31--45 (46\%), 46-60 (16\%),  and 61+ (4\%); 70\% were white, 16\% were black, 6\% were asian, and the remainder were mixed; and 31\% completed a bachelor's degree, 28\% completed high school, 13\% had an associate's degree, and the remaining completed graduate degrees. The MPQ survey found 65\% of respondents were risk-loving and 35\% were risk-averse. The HL survey found that 70\% of respondents were risk-averse, in direct contrast to the MPQ results. Participants generally trusted AI (55\% of subjects rated $\geq4$, 33\% rated 3, 12\% rated $\leq2$) and frequently used AI to help with their work (59\% responded $\geq4$, 18\% rated 3, 23\% rated $\leq2$).

\begin{figure}
    \centering
    \includegraphics[width=.4\textwidth]{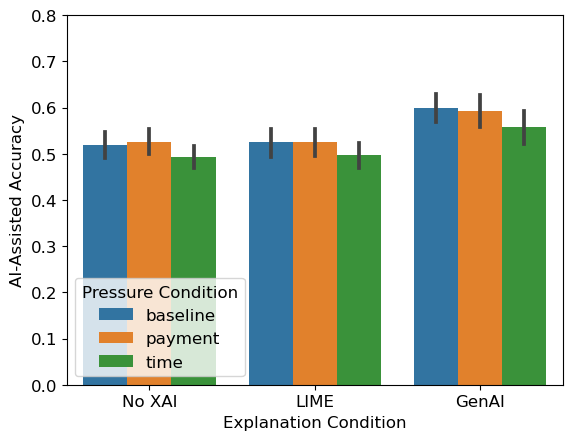}
    \caption{Accuracy after receiving AI advice.}
    \label{fig:accuracy}
\end{figure}

Figure \ref{fig:accuracy} summarizes accuracy after receiving AI advice across all conditions.  We note that \chatgpt advice increases task accuracy significantly over \noxai ($p<.001$) and \lime ($p<.001$);  time pressure slightly decreases accuracy, mirroring \citet{time-pressure}'s limited findings of how time pressure influences AI advice-taking.

\paragraph{Pressure via monetary loss has varying effects on AI advice usage, and time pressure lowers appropriate AI advice usage (H1).} 
Figure~\ref{fig:rair-rsr} summarizes the difference in relative positive AI- and self-reliance (respectively, RAIR on the left and RSR on the right), across conditions. 
We observe varied results with the payment pressure condition. When a natural language explanation is given,  RAIR drops by 5\% on average;  RSR decreases in both no-XAI (6\%) and GenAI conditions (8\%).
Time pressure has the strongest influence: RAIR and RSR decrease in \lime and drop significantly in \chatgpt ($p<.01$, $p<.01$, respectively). Participants completed these tasks faster, averaging 20 seconds per task compared to the 30 seconds used in the baseline and payment conditions, despite being given 30 seconds to complete each task.  Hence we attribute the performance drop in the time pressure condition to rushing.

We build generalized linear mixed models (GLMM) with Gaussian distributions to investigate how pressure and XAI conditions influence observed RAIR and RSR differences (Appendix \ref{sec:regression}). All dependent variables are log-transformed, and we consider an effect to be credible if the 97\% highest density interval (HDI) excluded zero. 

Neither pressure nor XAI support a clear directional effect for RAIR, but both variables are associated with reduced RSR ($\beta=-0.17$, HDI $[-0.33, -0.02]$; $\beta=-0.24$, HDI $[-0.34, -0.07]$). Confirming our observations that time pressure lowers overall accuracy, the models found pressure condition to have a credible negative association with accuracy ($\beta=-0.03$, HDI $[-.06, -.01]$) .

Overall, we find that payment pressure has an observable negative effect on appropriate AI advice reliance and overall accuracy, rejecting our hypothesis that performance pressure will improve appropriate AI reliance. The influence is most pronounced in the \chatgpt condition. The time pressure condition appears to have stressed participants out,  influencing them to rely more on the AI advice, resulting in lower RAIR, RSR, and overall accuracy.

\paragraph{Risk aversion scores have credible negative associations with RAIR but not RSR. AI advisor trust has overall negative associations with appropriate AI reliance. (H2).}
Our GLMM models find both risk aversion measures, MPQ and the Holt-Laury survey, are associated with reduced RAIR ($\beta=-0.33$, HDI $[-0.61, -0.05]$; $\beta=-0.39$, HDI $[-0.68, -0.10]$), but not RSR, despite reporting opposing results: the MPQ personality survey rated participants as risk-loving whereas the Holt-Laury survey rated the same participants as risk-averse, and vice versa. Neither risk score had a credible association with RSR and overall accuracy.
We attribute this result to the minimal influence of payment pressure on the dependent variables.
Participants' trust in the AI advisor had a credible \emph{negative} association with RAIR ($\beta=-0.89$, HDI $[-1.25, -0.43]$) but a \emph{positive} association with RSR ($\beta=0.74$, HDI $[0.33, 1.13]$). 
Neither risk nor trust measures had reliable associations with overall accuracy. From the postsurvey: 10 participants stated they relied on the AI after realizing the first few judgments aligned with the AI advice, indicating trust in the AI advisor, and 7 participants stated they guessed or followed their gut feeling.

\paragraph{Payment pressure increases decision making time (H3).}
We observe different decision-making behavior when users are operating under payment pressure despite its weak influence measured through RAIR and RSR.
In the payment pressure condition, participants on average spent 40\% longer making the initial decision and 10\% longer considering AI advice before making the final decision, as compared to the baseline pressure condition. This suggests payment pressure could act as a cognitive forcing function, encouraging more analysis before making a decision. Despite the increased time spent on tasks, overall performance was largely unaffected, with a minor increase when no XAI is present and a minor decrease when \chatgpt is available (Figure~\ref{fig:accuracy}).

\paragraph{XAI can mitigate negative effects of pressure (H4).}
In contrast to \citet{10.1145/3581641.3584066}, whose work found that LIME improved RAIR and had a minimal effect on RSR, our  results show LIME decreases RAIR compared to no-XAI and reduces RSR in the time pressure condition. Yet \lime synergizes well with payment pressure, improving RAIR over the baseline and maintaining RSR. The RSR change aligns with two participants who stated they relied on the AI significantly and entirely during the time conditions. GenAI explanations improved RAIR and RSR in the baseline condition, but pressure canceled out this effect. However, GenAI marginally improved overall accuracy compared to \noxai and \lime, indicating the potential for GenAI-style XAI in challenging tasks.

\subsection{How did participants judge reviews?}\label{sec:how-judge}
Our postsurvey asked participants to describe how they determined whether a review was genuine or deceptive. 
Several subjects admitted to guessing because they could not determine any relevant features indicating review quality (genuine or deceptive).
Twenty-four subjects checked for grammar, typos, and punctuation, and 23 examined the specifity of the reviews. 
Eight subjects ``tried to feel human emotions'', suggesting they associated deceptive reviews with being algorithmically generated, which describes algorithmic aversion for subjective tasks \citep{doi:10.1177/0022243719851788}.
Two subjects stated they were determined not to ``lose some bonus'', indicating that the payment pressure condition affected their behavior.

\section{Discussion} \label{sec:discussion}
This work investigated how environmental pressure, combined with XAI, influences the way crowdworkers complete an AI-assisted decision making task. A large body of literature has focused on improving AI reliance isolated from environmental factors such as pressure, but AI-assisted decision making often occurs under pressure. We demonstrated two simple methods for inducing  pressure through payment framed as a loss and limiting time.

\subsection{AI advice reliance under pressure}
There are many types of environmental pressures that an AI-assisted decision maker can be under, and interaction effects can vary widely. When looking at payment pressure,
our findings show a subtle effect of how AI advice reliance changes.
Without explanations, the pressure conditions had minimal influence on RAIR but negative influence on RSR. With \lime, RAIR improved and RSR remained the same with payment pressure, but RAIR and RSR both decreased with time pressure. With \chatgpt, RAIR and RSR consistently decreased.

The RAIR decrease in \lime in the baseline pressure condition compared to \noxai suggests \lime decreased trust in the AI advice.\footnote{ Disentangling AI trust and AI reliance behavior is its own challenge \citep{scharowski2022trust,10.1145/3630106.3658901}} It associated individual words with a prediction that may not have made sense (e.g., ``is'' is associated with being genuine; \autoref{fig:judge-lime}), but users reported associating typos, excessive punctuation, and grammar with their prediction (\S\ref{sec:how-judge}). Payment pressure appears to mitigate the negative effects of LIME, improving RAIR over the LIME baseline (but not over the \noxai baseline) without changing RSR. This result points toward the potential for XAI techniques in the same style as LIME to help mitigate or further improve AI-assisted decision making behavior on tasks where humans are close to random.

\chatgpt has the worst interaction with pressure. RAIR decreases to the level of \noxai in the payment condition and to the level of \lime in the time condition.
We suspect that the payment pressure increased skepticism in AI advice, since natural language explanations can appear  generic (e.g., ``The review is deceptive because it overly praises the hotel without mentioning any potential downsides''). The time spent considering AI advice increased over the baseline, which can be interpreted as payment pressure influencing people to take more care in their final decision, or reducing their trust in the AI advice in a similar manner to how narcissism has a negative relationship with advice taking \citep{KAUSEL201533,OREILLY2021110280}.

One XAI method can be more effective than another depending on the task or stakeholders \citep{jiang2022needs}, and our results show that this is the case even under different pressure conditions. Furthermore, the negative effect of pressure can override the potential benefits of XAI, as seen in the \chatgpt RAIR decrease.

Time pressure has a largely consistent, 
negative impact on RAIR, RSR, and overall performance. The behavior change mirrors the phenomenon of ``choking,'' where a human experiences performance decline in critical high pressure settings, often attributed to anxiety. Participants were anxious about completing the tasks within the 30 second time limit, and rushed  to complete them faster.
As a result, they relied heavily on the AI advice. One subject stated: ``For the timed cases, I ended up relying on the AI's decision and just went with it. However, for the other parts, I looked out for typos and missing punctuations to detect human reviews.''

\citet{time-pressure} observed that slowing down AI overreliers in higher time pressure environments could improve their performance, and since our participants had plenty of time to spare, we expect such a strategy would be helpful in this setting. Emotional regulation strategies can be effective in improving performance when anxiety is high \citep{balk2013coping}, which can help alleviate anxiety over running out of time (and hence the possibility of earning a bonus).
Methods from the distraction model may also help: if choking occurs due to cognitive overload (rather than anxiety), an attention shift could refocus the user to pay attention to relevant information and avoid choking \citep{hardy2001effect,mullen2005effects,eysenck2012anxiety,nieuwenhuys2012anxiety}. For example, adding a third color to \lime explanations, unrelated to the prediction, could be a distractor.

\subsection{Practical Advice}

A motivating use case for this work is AI-assisted crowdsourced annotation and data elicitation projects in NLP research.  In summary, we recommend that GenAI-based explanations be explored for benefits to appropriate AI advice reliance and accuracy. For projects interested in exploring how annotations or elicited data change under different pressure environments, e.g., users may prefer curt dialogue interactions in high pressure environments but elaborate responses otherwise, care should be taken to ensure confirmation bias does not increase AI trust throughout the course of the project.
Performance pressure may be a useful tool, but we did not find consistent benefits.

\section{Future Directions}

Our work has illustrated how environmental stressors and XAI can alter AI-assisted decision making behavior. 
The mixed effectiveness of incorporating XAI under pressures reinforces the findings of \citet{jiang2022needs} and \citet{time-pressure} where AI advice or XAI are only effective for some people, some environments, and at certain times.
The challenges we observed in determining significant predictive factors of AI advice taking behavior also reflect the difficulty in modeling appropriate AI advice reliance. 
This calls for building adaptive environments that can be personalized for the user and task to provide appropriate XAI methods, cognitive forcing functions, and other AI advice interventions. 
Such environments could improve AI-assisted decision making and help elicit more diverse data for training robust NLP models.
However, in order to develop these environments, we need a deeper understanding of how environmental factors interact with user characteristics and different XAI methods.
Future work should explore other pressures such as competition and multi-tasking, alongside XAI and measures of user personality and behavior, e.g., how they cope with stress.
Additional work investigating the influence of performance pressure on tasks where humans have higher than random chance performance should also be investigated, as the literature suggests performance pressure can improve performance, despite our results suggesting minimal influence.

\section{Conclusion}
AI assistance is already used in the real world for tasks with high and low risk, and designing adaptive AI assistants that are domain-specific requires understanding the different factors influencing how humans use AI advice.
We investigated how  pressure influences the use of AI advice.
Using deceptive review classification, crowdworkers with little expertise or personal motivation requirements, and two different XAI techniques, we observed complex effects on how pressure influences AI advice usage.
Pressure and XAI interactions could both improve and decrease appropriate AI advice.
While performance pressure had minor effects, time pressure had a strong negative effect.
Our results contribute to the body of literature investigating AI-assisted decision making.
We note the relevance of these findings in AI annotation projects in particular; our work motivates continued  research on the effects of pressure on AI assistance in varied environments while taking into account individual differences. 

\section*{Limitations}
Our experiments in manipulating environmental pressure used an inherently low-stakes, challenging task completed by crowdworkers. As a relatively young area of inquiry, it would be important to investigate how pressure and XAI influence changes between laypeople and experts. We expect different behavior because expertise can influence information seeking behavior \cite{expertise-information-seeking} and buffer stress responses \citep{10.1093/oxfordhb/9780198795872.013.22}.

The literature has yet to find consensus on how people behave in AI-assisted decision making. Our \lime-baseline results contrast with those of \citet{10.1145/3581641.3584066}, despite using a similar interface and pool of crowdworkers. Understanding AI-assisted decision making is a complex and challenging endeavor, and the rapid adoption of AI assistance further motivates research in this area.

\section*{Ethical Considerations}
Our work investigated how pressure and explainable AI influence AI-assisted decision making, and our results show how trust in the AI system plays an important role in agreeing with AI advice. A malicious actor could design a system to increase AI trust in order to persuade others to agree with AI advice, against their better judgment, and overriding beneficial influences of pressure that persuade people to be more skeptical and careful. For example, 10 of our participants stated they found the AI advice agreed with their judgments for the first few instances, leading them to rely on the AI advice more at later stages. Imposing a time limit on the decision could further convince people to rely on the AI advice because relying on the AI is an easy method to cope with time stress.

\section*{Acknowledgements}
This research used statistical consulting resources provided by the Center for Statistics and the Social Sciences, University of Washington. We thank Dr. Thomas Frye for valuable insights in leveraging methods from behavioral economics.

\bibliography{custom}

\appendix
\section{Methods} 
\label{app:prompt-chatgpt}
The natural language explanations for the \chatgpt XAI condition were generated using ChatGPT with default settings on Jan 20, 2025, with the prefix prompt: 
In less than 50 words, explain why the following review is a [genuine|deceptive] review.
\begin{figure*}[!th]
    \centering
     \begin{subfigure}[b]{0.48\textwidth}
         \centering
         \includegraphics[width=\textwidth]{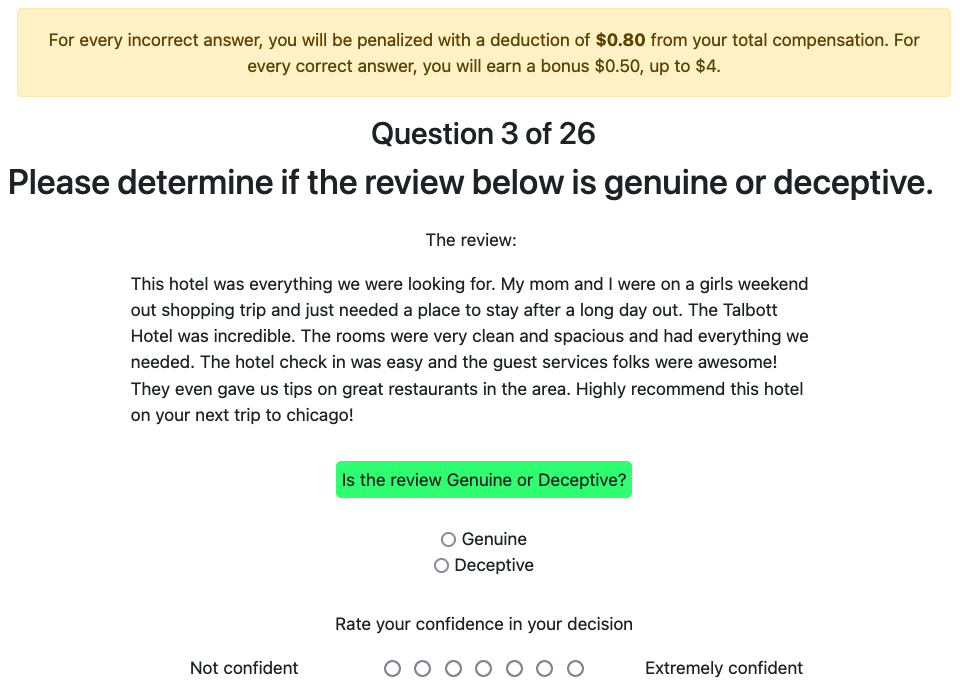}
         \caption{The initial decision}
         \label{fig:judge-1}
     \end{subfigure}
     \hfill
     \begin{subfigure}[b]{0.48\textwidth}
         \centering
         \includegraphics[width=\textwidth]{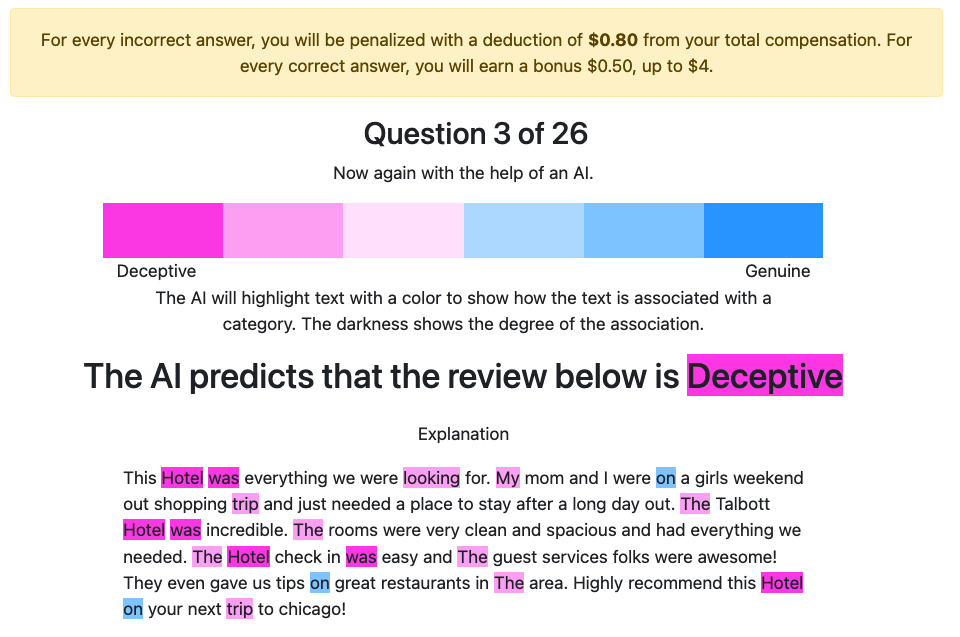}
         \caption{Second page of the judgment task displaying AI advice alongside \lime explanation}
         \label{fig:judge-lime}
     \end{subfigure}
     \hfill
        \caption{The user interface in payment-\lime condition. The user first encounters the review in (a) and makes a judgment. Then they receive AI advice (b) in the form of a model prediction and need to make their judgment again. }
        \label{fig:user-interface-xai}
\end{figure*}

\subsection{RAIR and RSR}\label{rair-rsr-math}
Relative AI reliance (RAIR) and relative self-reliance (RSR) are defined by \citet{10.1145/3581641.3584066}, which we reproduce below for reference.
Subjects complete a prediction task with $N$ instances $x_i\in X$ with ground truth labels $y_i \in Y$.
RAIR is the ratio of cases where the human correctly changes their decision to follow AI advice.
\begin{equation}\label{math:rair}
RAIR = \frac{\sum_{i=0}^{N}CAIR_i}{\sum_{i=0}^{N}CA_i}
\end{equation}
where 
 
\begin{enumerate}
    \item $CAIR$: 1 if the initial judgment disagrees with the ground truth, the AI advice is correct, and the final judgment agrees with the ground truth; 0 otherwise
    \item $CA$: 1 if the initial judgment is incorrect and the AI advice is correct, regardless of the final judgment; 0 otherwise 
\end{enumerate}
RSR is the ratio of cases where the human correctly disregards incorrect AI advice.

\begin{equation}\label{math:rsr}
RAIR = \frac{\sum_{i=0}^{N}CSR_i}{\sum_{i=0}^{N}IA_i}
\end{equation}
\begin{enumerate}
    \item $CSR$: 1 if the initial judgment agrees is correct, the AI advice is incorrect, and the final judgment is correct; 0 otherwise
    \item $IA$: 1 if the initial judgment is correct and the AI advice is incorrect; 0 otherwise
\end{enumerate}

\subsection{Presurvey} 
\label{presurvey}
The presurvey contained demographics questions and the 10-item IPIP adaptation of the \citet{mpq} Multiple Personality Questionnaire (MPQ).
Questions:
\begin{enumerate}
    \item Indicate your age range
    \item Indicate your race
    \item Indicate your gender
    \item What is your highest level of education completed?
    \item What is your native language?
    \item Describe your proficiency in other languages
    \item How familiar are you with the task of deciding whether a review is genuine or deceptive? (5-point Likert)
    \item Rate your level of agreement with the following statements (5-point Likert): (1) I trust artificial intelligence (AI); (2) I use AI to help me with my work
\end{enumerate}

MPQ consists of 10 questions answered on a 4-point Likert scale. Subjects are asked to rate their level of agreement with each statement.
\label{mpq-questions}
\begin{enumerate}
\item I would never go hang gliding or bungee jumping.
\item I would never make a high-risk investment.
\item I avoid dangerous situations.
\item I seek danger.
\item I am willing to try anything once.
\item I do dangerous things.
\item I enjoy being reckless
\item I seek adventure
\item I take risks
\item I do crazy things
\end{enumerate}

\subsection{Postsurvey}\label{postsurvey}
Postsurvey questions:
\begin{enumerate}
    \item Trust in AI Advisor: Rate your level of agreement with the following statements on a 7-point Likert scale. 1) I think I can trust the AI Advisor; 2) The AI advisor can be trusted to provide reliable support; 3) I trust the AI advisor to keep my best interests in mind; 4) In my opinion, the AI advisor is trustworthy
    \item HL survey
    \item Describe how you determined whether a review was genuine or deceptive.
\end{enumerate}
The Trust in AI Advisor questions were sourced from \citet{10.1145/3581641.3584066}, who sourced them from \citealp{doi:10.1177/002224299005400306,doi:10.1177/002224299706100203,doi:10.1177/002224299405800201} and \citealp{10.5555/2017181.2017185}.

\subsubsection{Holt-Laury Survey}\label{hl-survey}
The Holt-Laury survey measures how risk averse a respondent is by asking them to make a decision between pairs of gambles with a ``safe'' and ``risky'' choice. For example, in the first question, the participant can select the safe choice of having 1/10 chance to earn \$2 and 9/10 chance of earning \$1.60, or the risky choice of having 1/10 chance to earn \$3.85 and 9/10 chance of earning \$0.10. One of the choices is randomly selected and the gamble played to determine the bonus to the respondent.

\begin{table}[!htp]\centering
\scriptsize
\begin{tabular}{lrr}\toprule
Safe choice &Risky choice \\\midrule
1/10 of \$2.00, 9/10 of \$1.60 &1/10 of \$3.85, 9/10 of \$0.10 \\
2/10 of \$2.00, 8/10 of \$1.60 &2/10 of \$3.85, 8/10 of \$0.10 \\
3/10 of \$2.00, 7/10 of \$1.60 &3/10 of \$3.85, 7/10 of \$0.10 \\
4/10 of \$2.00, 6/10 of \$1.60 &4/10 of \$3.85, 6/10 of \$0.10 \\
5/10 of \$2.00, 5/10 of \$1.60 &5/10 of \$3.85, 5/10 of \$0.10 \\
6/10 of \$2.00, 4/10 of \$1.60 &6/10 of \$3.85, 4/10 of \$0.10 \\
7/10 of \$2.00, 3/10 of \$1.60 &7/10 of \$3.85, 3/10 of \$0.10 \\
8/10 of \$2.00, 2/10 of \$1.60 &8/10 of \$3.85, 2/10 of \$0.10 \\
9/10 of \$2.00, 1/10 of \$1.60 &9/10 of \$3.85, 1/10 of \$0.10 \\
10/10 of \$2.00, 0/10 of \$1.60 &10/10 of \$3.85, 0/10 of \$0.10 \\
\bottomrule
\end{tabular}
\caption{Holt-Laury survey}\label{tab: }
\end{table}

\section{Results}
 
\begin{table}[!htp]\centering
    \begin{tabular}{llll}
    \toprule
    XAI                                        & Pressure & Pre-Advice & Post-Advice \\ \midrule
    \multicolumn{1}{c}{\multirow{4}{*}{NoXAI}} & Baseline & 0.52       & 0.55        \\
    \multicolumn{1}{c}{}                       & Payment  & 0.54       & 0.55        \\
    \multicolumn{1}{c}{}                       & Time     & 0.51       & 0.52        \\
    \multicolumn{1}{c}{}                       & Overall  & 0.52       & 0.54        \\ \midrule
    \multirow{4}{*}{LIME}                      & Baseline & 0.52       & 0.55        \\
                                               & Payment  & 0.54       & 0.55        \\
                                               & Time     & 0.51       & 0.52        \\
                                               & Overall  & 0.52       & 0.54        \\ \midrule
    \multirow{4}{*}{GenAI}                     & Baseline & 0.52       & 0.55        \\
                                               & Payment  & 0.54       & 0.55        \\
                                               & Time     & 0.51       & 0.52        \\
                                               & Overall  & 0.52       & 0.54        \\ \bottomrule
    \end{tabular}
    \caption{Accuracy before and after receiving AI advice. Accuracy marginally increased after receiving AI advice.}
    \label{tab:pressure_accuracy}
\end{table}

\subsection{Significance Tests}\label{sec:sigtests}

A Shapiro test for normality indicates our data is not normal. To test for significance, we use the Friedman test with Wilcoxon post-hoc tests for our within-subjects (pressure condition) data and the Kruskal-Wallis and Dunn post-hoc tests for our between-subjects (XAI condition) data. We report uncorrected and Holm-Bonferroni corrected p-values when significant.

\paragraph{Pressure condition tests.} For RAIR, when \chatgpt was present, the difference between baseline and time pressure tasks had p-values of $p=.009$, corrected to $p=.028$. For RSR, when \chatgpt was present, there was a significant difference between baseline and payment pressures ($p=.008$, corrected  $p=.016$), and baseline and time pressures ($p=.003$, corrected $p=.009$).
\paragraph{XAI condition tests.} For RSR, GenAI significantly differed from both the baseline \noxai ($p=8\mathrm{e}{-6}$, corrected $p=2.4\mathrm{e}{-5}$) and \lime ($p=3.4\mathrm{e}{-4}$, corrected $p=.0007$). Similarly, for overall task accuracy, \chatgpt significantly differed from \noxai ($p=1.7\mathrm{e}{-7}$, corrected $p=5.0\mathrm{e}{-7}$) and \lime ($p=7\mathrm{e}{-7}$, corrected $p=1\mathrm{e}{-6}$).

\subsection{Regression Tables}\label{sec:regression}
The generalized linear mixed models using a Gaussian distribution and log-transformed dependent variable results are detailed below. Values in parentheses were transformed back to the original scale.

\label{regressiont-rair}

    \begin{table*}[ht]
        \centering\tiny
    \begin{tabular}{llrllrrrrr}
\toprule
DV: RAIR & Mean ($\beta$) & SD & HDI 3\% & HDI 97\% & MCSE Mean & MCSE SD & ESS Bulk & ESS Tail & $\hat{R}$ \\
\midrule
Sigma & 2.101 & 0.050 & 2.005 & 2.192 & 0.000 & 0.001 & 11390.000 & 6135.000 & 1.000 \\
Intercept & -1.523 & 0.225 & -1.943 & -1.095 & 0.002 & 0.003 & 12491.000 & 6218.000 & 1.000 \\
\midrule
Pressure & -0.107 (0.899) & 0.085 & -0.269 (0.764) & 0.049 (1.05) & 0.001 & 0.001 & 11686.000 & 5580.000 & 1.000 \\
Explanation & -0.07 (0.932) & 0.088 & -0.232 (0.793) & 0.097 (1.102) & 0.001 & 0.001 & 11010.000 & 6630.000 & 1.000 \\
MPQ\textasteriskcentered{} & -0.332 (0.717) & 0.151 & -0.608 (0.544) & -0.047 (0.954) & 0.001 & 0.002 & 11254.000 & 6139.000 & 1.000 \\
Holt-Laury\textasteriskcentered{} & -0.391 (0.676) & 0.155 & -0.682 (0.506) & -0.095 (0.909) & 0.001 & 0.002 & 12901.000 & 6782.000 & 1.000 \\
Advisor Trust\textasteriskcentered{} & -0.858 (0.424) & 0.218 & -1.248 (0.287) & -0.432 (0.649) & 0.002 & 0.002 & 11377.000 & 6740.000 & 1.000 \\
Trust in AI & -0.066 (0.936) & 0.064 & -0.188 (0.829) & 0.052 (1.053) & 0.001 & 0.001 & 10956.000 & 6341.000 & 1.000 \\
Age & -0.121 (0.886) & 0.071 & -0.249 (0.78) & 0.014 (1.014) & 0.001 & 0.001 & 11665.000 & 6379.000 & 1.000 \\
Race & 0.015 (1.015) & 0.031 & -0.042 (0.959) & 0.073 (1.076) & 0.000 & 0.000 & 11200.000 & 6051.000 & 1.000 \\
Gender & 0.167 (1.182) & 0.122 & -0.061 (0.941) & 0.392 (1.48) & 0.001 & 0.001 & 11282.000 & 6049.000 & 1.000 \\
Education\textasteriskcentered{} & 0.072 (1.075) & 0.027 & 0.02 (1.02) & 0.12 (1.127) & 0.000 & 0.000 & 11259.000 & 5936.000 & 1.000 \\
\bottomrule
\end{tabular}

    \caption{GLMM results for RAIR. * indicates credible effect (97\% HDI excludes 1). Mean and HDI values in parentheses have been transformed back to the original scale.}
    \label{tab:reg-rair-2}
    \end{table*}

\label{regression-rsr}

    \begin{table*}[ht]
        \centering\tiny
    \begin{tabular}{llrllrrrrr}
\toprule
 DV: RSR & Mean ($\beta$) & SD & HDI 3\% & HDI 97\% & MCSE Mean & MCSE SD & ESS Bulk & ESS Tail & $\hat{R}$ \\
\midrule
Sigma & 2.027 & 0.048 & 1.936 & 2.116 & 0.000 & 0.001 & 12155.000 & 6525.000 & 1.000 \\
Intercept & -1.394 & 0.214 & -1.802 & -1.003 & 0.002 & 0.003 & 12251.000 & 6168.000 & 1.000 \\
\midrule
Pressure\textasteriskcentered{} & -0.169 (0.845) & 0.084 & -0.331 (0.718) & -0.015 (0.985) & 0.001 & 0.001 & 12549.000 & 6463.000 & 1.000 \\
Explanation\textasteriskcentered{} & -0.24 (0.787) & 0.085 & -0.397 (0.672) & -0.074 (0.929) & 0.001 & 0.001 & 12539.000 & 6400.000 & 1.000 \\
MPQ & 0.223 (1.25) & 0.143 & -0.038 (0.963) & 0.491 (1.634) & 0.001 & 0.002 & 11592.000 & 6475.000 & 1.000 \\
Holt-Laury & 0.111 (1.117) & 0.147 & -0.158 (0.854) & 0.391 (1.478) & 0.001 & 0.002 & 12765.000 & 6687.000 & 1.000 \\
Advisor Trust\textasteriskcentered{} & 0.738 (2.092) & 0.211 & 0.33 (1.391) & 1.125 (3.08) & 0.002 & 0.002 & 10998.000 & 6235.000 & 1.000 \\
Trust in AI & 0.029 (1.029) & 0.060 & -0.079 (0.924) & 0.147 (1.158) & 0.001 & 0.001 & 11932.000 & 6689.000 & 1.000 \\
Age & 0.053 (1.054) & 0.069 & -0.076 (0.927) & 0.182 (1.2) & 0.001 & 0.001 & 11936.000 & 6174.000 & 1.000 \\
Race & 0.0 (1.0) & 0.029 & -0.055 (0.946) & 0.053 (1.054) & 0.000 & 0.000 & 12330.000 & 6358.000 & 1.000 \\
Gender\textasteriskcentered{} & -0.23 (0.795) & 0.115 & -0.447 (0.64) & -0.018 (0.982) & 0.001 & 0.001 & 13571.000 & 6405.000 & 1.000 \\
Education & -0.019 (0.981) & 0.025 & -0.067 (0.935) & 0.029 (1.029) & 0.000 & 0.000 & 10931.000 & 6202.000 & 1.000 \\
\bottomrule
\end{tabular}

    \caption{GLMM results for RSR. * indicates credible effect (97\% HDI excludes 1). Mean and HDI values in parentheses have been transformed back to the original scale.}
    \label{tab:reg-rsr-3}
    \end{table*}

\newpage
\label{regression-accuracy}

    \begin{table*}[ht]
        \centering\tiny
    \begin{tabular}{llrllrrrrr}
\toprule
 DV: Overall Accuracy & Mean ($\beta$) & SD & HDI 3\% & HDI 97\% & MCSE Mean & MCSE SD & ESS Bulk & ESS Tail & $\hat{R}$ \\
\midrule
Sigma & 0.317 & 0.008 & 0.303 & 0.331 & 0.000 & 0.000 & 11965.000 & 6325.000 & 1.000 \\
Intercept & -0.532 & 0.034 & -0.598 & -0.472 & 0.000 & 0.000 & 11254.000 & 6552.000 & 1.000 \\
\midrule
Pressure\textasteriskcentered{} & -0.032 (0.969) & 0.013 & -0.056 (0.946) & -0.008 (0.992) & 0.000 & 0.000 & 12719.000 & 5529.000 & 1.000 \\
Explanation\textasteriskcentered{} & -0.06 (0.942) & 0.013 & -0.084 (0.919) & -0.034 (0.967) & 0.000 & 0.000 & 11629.000 & 6508.000 & 1.000 \\
MPQ & -0.011 (0.989) & 0.023 & -0.052 (0.949) & 0.033 (1.034) & 0.000 & 0.000 & 12045.000 & 6433.000 & 1.000 \\
Holt-Laury & -0.018 (0.982) & 0.023 & -0.06 (0.942) & 0.025 (1.025) & 0.000 & 0.000 & 11983.000 & 6352.000 & 1.000 \\
Advisor Trust & -0.013 (0.987) & 0.034 & -0.077 (0.926) & 0.05 (1.051) & 0.000 & 0.000 & 11390.000 & 6084.000 & 1.000 \\
Trust in AI & 0.002 (1.002) & 0.009 & -0.017 (0.983) & 0.019 (1.019) & 0.000 & 0.000 & 11321.000 & 6503.000 & 1.000 \\
Age & -0.009 (0.991) & 0.011 & -0.029 (0.971) & 0.012 (1.012) & 0.000 & 0.000 & 11287.000 & 6351.000 & 1.000 \\
Race & -0.003 (0.997) & 0.005 & -0.012 (0.988) & 0.006 (1.006) & 0.000 & 0.000 & 12220.000 & 6107.000 & 1.000 \\
Gender & -0.025 (0.975) & 0.018 & -0.059 (0.943) & 0.011 (1.011) & 0.000 & 0.000 & 12496.000 & 6420.000 & 1.000 \\
Education & 0.007 (1.007) & 0.004 & -0.0 (1.0) & 0.015 (1.015) & 0.000 & 0.000 & 11597.000 & 6548.000 & 1.000 \\
\bottomrule
\end{tabular}

    \caption{GLMM results for overall accuracy. * indicates credible effect (97\% HDI excludes 1). Mean and HDI values in parentheses have been transformed back to the original scale.}
    \label{tab:reg-ai_assisted_accuracy_mean-4}
    \end{table*}

\pagebreak\newpage\clearpage

\newpage
\section{Licenses}
The Deceptive Opinion Spam Corpus v1.4 \citep{ott-etal-2011-finding,ott-etal-2013-negative} is licensed under Creative Commons Attribution-NonCommercial-ShareAlike 3.0 Unported License. The outputs from ChatGPT, a large language model from OpenAI, are copyright free. 

\section{AI Use}
ChatGPT was used to help with data transformation.

\end{document}